\documentclass{webofc}
\usepackage[varg]{txfonts}

\begin{document}

\title{Locating the special point of hybrid neutron stars}

\author{\firstname{Mateusz} \lastname{Cierniak}\inst{1}\fnsep\thanks{\email{mateusz.cierniak@uwr.edu.pl}} \and
        \firstname{David} \lastname{Blaschke}\inst{1,2,3}\fnsep\thanks{\email{david.blaschke@uwr.edu.pl}}
}

\institute{
            University of Wroclaw, 50-137 Wroclaw, Poland
\and
           Bogoliubov Laboratory of Theoretical Physics, JINR Dubna, 141980 Dubna, Russia
\and
           National Research Nuclear University (MEPhI), 115409 Moscow, Russia
          }
          
\abstract{%
The special point is a feature unique to models of hybrid neutron stars. 
It represents a location on their mass--radius sequences that is insensitive to the phase transition density.
We consider hybrid neutron stars with a core of deconfined quark matter that obeys a constant--sound--speed (CSS) equation of state model and provide a fit formula for the coordinates of the special point as functions of 
the squared sound speed ($c_s^2$) and pressure scale ($A$) parameters.
Using the special point mass as  a proxy for the maximum mass of the hybrid stars we derive limits for the CSS model parameters based on the recent NICER constraint on mass and radius of pulsar PSR J0740+6620, 
$0.36 < {c^2_s}_{\rm min} < 0.43$ and $80<A [{\rm MeV/fm}^3]<160$.
The upper limit for the maximum mass of hybrid stars depends on the upper limit for $c_s^2$ so that 
choosing $c^2_{s,max} = 0.6$ results in $M_{\rm max}<2.7~M_\odot$, within the mass range of GW190814.
}

\maketitle

\section{Introduction}
\label{intro}

The recent observation of gravitational waves originating from the inspiral of a binary system of neutron stars has resulted in a significant step forward in the field of nuclear astrophysics. 
The event, labeled GW170817 \cite{LIGOScientific:2017vwq}, along with a subsequent observation of the resulting kilonova \cite{LIGOScientific:2017ync} signaled the beginning of the era of multi-messenger astronomy. 
As a direct result, for the first time, it became possible to determine limits on the neutron star tidal deformability, and due to its connection with the equation of state (EoS)  of dense neutron star matter \cite{Hinderer:2009ca} also to obtain limits on the 
masses and radii of the merging stars \cite{Annala:2017llu, Bauswein:2017vtn}.
In addition, this event has reignited the discussion on signals of quark- matter in the cores of neutron stars \cite{Bauswein:2018bma,Bauswein:2020aag}.

The possibility of finding this rare state of matter inside neutron stars provides a unique opportunity of accessing the cold and dense sector of the Quantum Chromodynamics (QCD) phase diagram, which remains out of reach for first--principle lattice QCD calculations.
A large body of work exists on the possible states of cold, strongly interacting matter and their relation to neutron star properties 
(for recent reviews, see \cite{Baym:2017whm,Blaschke:2018mqw} and references therein), which were triggered by the observation of pulsars with masses around $2~M_\odot$ \cite{Demorest:2010bx,Antoniadis:2013pzd,NANOGrav:2019jur,Fonseca:2021wxt}.
But without a radius measurement at this high mass, there are still too many EoS scenarios possible, with and without a deconfinement phase transition.


Efforts are being made to solve the problem of a reliable radius measurement by employing the X-ray timing method (cf. \cite{Bogdanov:2006zd} and references therein). 
For that purpose, the NASA NICER mission was deployed to the International Space Station in June 2017 and has since then gathered luminosity data from a few millisecond pulsars. Results have been published for two objects,:
the heaviest known pulsar with a precisely determined mass, PSR J0740+6620 \cite{Miller:2021qha,Riley:2021pdl}, and an intermediate mass pulsar, PSR J0030+0451 \cite{Miller:2019cac, Riley:2019yda}.
The latter only marginally agrees with the analysis of GW170817 \cite{LIGOScientific:2018cki} in the same mass range, 
resulting in a narrow region of possible radii for intermediate-mass neutron stars \cite{Capano:2019eae}, 
or perhaps hint at a signal of the third family of compact stars \cite{Gerlach:1968zz}
which would give rise to the phenomenon of mass twin stars \cite{Glendenning:1998ag} that could occur in different mass 
ranges \cite{Christian:2017jni}, also in the mass range of GW170817, see  \cite{Blaschke:2018mqw} and references therein.

The high-mass pulsar PSR J0740+6620 is observed to have a rather large radius, see figure~\ref{fig-1}.
This creates tension with the idea of purely hadronic neutron stars. 
Hadronic matter tends to soften at higher densities due to the onset of hyperons \cite{Chatterjee:2015pua} which creates the 
"hyperon puzzle" problem for nonrelativistic EoS models that are based on realistic nucleon-nucleon interactions like the APR EoS
\cite{Akmal:1998cf}. Adding a repulsive few-nucleon interaction in the form of a multi-pomeron potential solves the hyperon puzzle for such models \cite{Yamamoto:2015lwa,Yamamoto:2017wre} but this cannot explain the large radius for PSR J0740+6620.

While not completely excluded, the scenario of purely hadronic neutron star interiors is being increasingly disfavored by observation.
A possible alternative are multi--phase models, predicting a phase transition from purely hadronic neutron stars at low mass to intermediate and high mass hybrid stars with a core presenting exotic states of matter, for instance a quark--gluon plasma (QGP).
Such constructions are not completely free of the shortcomings of hadronic models, as the common convention is to construct a Maxwell--type first order phase transition from a stiffer hadronic equation of state (EoS) to a softer one, represented by the QGP.
Recently, a number of alternative approaches were proposed \cite{Masuda:2012ed,Baym:2017whm,Ayriyan:2021prr} which join a soft low-density hadronic EoS with stiff QGP models at high densities, thus easily fulfilling all observational constraints.

In this manuscript, we will restrict ourselves to the standard Maxwell construction of a phase  transition and make use of the novel insights due to the special point property of hybrid neutron star sequences \cite{Yudin:2014mla,Cierniak:2020eyh,Blaschke:2020vuy,Cierniak:2021knt} in order to determine the impact of the recent observations on the available parameter space of a class of constant--speed--of--sound (CSS) QGP models \cite{Alford:2013aca}, which were shown to provide an accurate description of color--superconducting quark matter \cite{Zdunik:2012dj,Antic:2021zbn}.

The manuscript is organized as follows. 
After introducing the two-phase hybrid EoS model for dense matter in the next section, we present results of an extensive scan of the possible locations of the special point in the mass-radius plane in section 3. In section 4, we introduce a fit formula that relates the coordinates of the special point to the CSS model parameters $A$ and $c_s^2$ and obtain as the main result of this contribution the mapping of the new NICER mass-radius constraint for PSR J0740+6620 onto the $A - c_s^2$ plane. In section 5 we draw our Conclusions. 

\section{The two phase dense matter model}
\label{sec-1}

In the absence of a unified description of cold hadronic and quark matter with a phase transition at high densities, the hybrid star matter EoS must be constructed by matching separate EoS, comparing pressures derived using hadrons, $P_h(\mu)$, or deconfined quarks, $P_q(\mu)$, as degrees of freedom.
The EoS with the higher pressure is the one that is preferred by nature and describes the bulk thermodynamics. 
A crossing of both alternative EoS at the critical chemical potential $\mu_c$ marks a first-order phase transition fulfilling the condition $P_h(\mu_c)=P_q(\mu_c)$ for the Maxwell construction.

The hadronic EoS discussed in this study is chosen from the class of relativistic density-functional (RDF) models based on the 
DD2 parametrization \cite{Typel:2009sy} with excluded volume effects modelled according to \cite{Typel:2016srf}. 
Its inputs are fitted to the properties of finite nuclei and it predicts the EoS for nuclear matter at low densities up to and at nuclear saturation in accordance with the phenomenological data. Therefore, it should provide a reliable basis for extrapolations to supersaturation densities.

For the deconfined quark matter phase, the class of CSS quark matter models will be used. 
The EoS in the form of pressure as a function of the baryochemical potential reads (see also \cite{Cierniak:2020eyh} or the Appendix of \citep{Alford:2013aca}), 
\begin{equation} 
\label{css3}
    P(\mu)=A(\mu/\mu_0)^{1+c_s^{-2}}-B,
\end{equation}
where the model parameters $A$, $B$ and $c_s^2$ are constant and $\mu_0=1$ GeV.
The baryon density $n$ follows from the canonical relation
\begin{equation}
\label{eq:density}
    \frac{d P(\mu)}{d \mu}= n(\mu)
    =(1+c_s^{-2})\frac{A}{\mu_0}\left(\frac{\mu}{\mu_0}\right)^{c_s^{-2}}.
\end{equation}
Using the above, we arrive at the energy density
\begin{equation}
    \varepsilon=\mu n-P=B + c_s^{-2} A(\mu/\mu_0)^{1+c_s^{-2}}.
\end{equation}
The relation between pressure and energy density takes the form
\begin{equation} \label{css6}
   P = c_s^2  \varepsilon - (1+c_s^2)B.
\end{equation}
It has been shown in \cite{Zdunik:2012dj} that the above CSS model fits well the EoS of color superconducting quark matter in both, the 2SC and the CFL phases which were obtained from a self-consistent solution of the three-flavor NJL model in the mean field approximation \citep{Blaschke:2005uj,Klahn:2013kga}. 

\section{The special point}
\label{sec-2}

\begin{figure}[h]
\centering
\includegraphics[width=8cm]{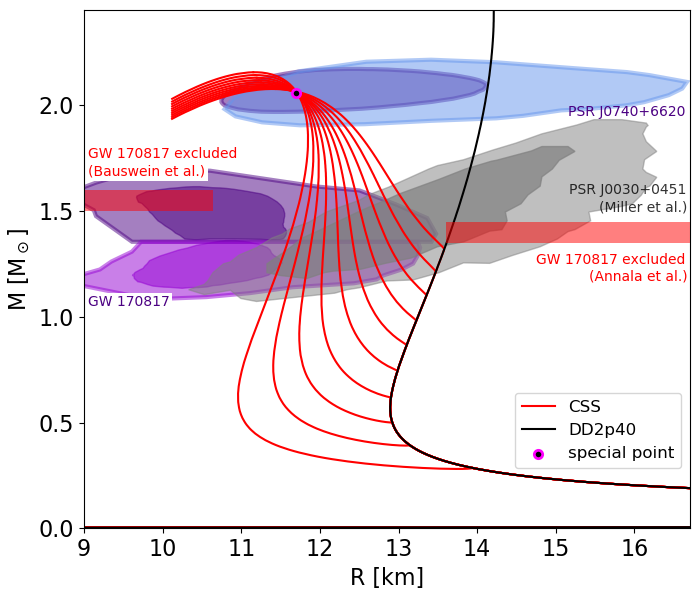}
\caption{Special point as the intersection of hybrid star sequences (red solid lines) on the neutron star mass radius diagram. 
The black line represents the hadronic EoS with excluded volume DD2p40, naming convention analogous to \cite{Typel:2016srf}. 
The intersection of the red lines with the black one mark the onset of CSS quark matter in the core, see also \cite{Cierniak:2021knt}.
Observational data taken from \cite{Miller:2021qha,Miller:2019cac,Riley:2021pdl,Riley:2019yda,LIGOScientific:2020zkf,Bauswein:2017vtn,Annala:2017llu}}
\label{fig-1}
\end{figure}

The special point of hybrid neutron stars is a property first described in \cite{Yudin:2014mla} and recently studied in \cite{Cierniak:2020eyh,Blaschke:2020vuy,Cierniak:2021knt}.
It is a characteristic feature of a broad range of two--phase models (see, e.g., \cite{Blaschke:2010vd,Kaltenborn:2017hus,Blaschke:2020qqj}), presenting itself as a stationary point in the mass--radius 
diagram for hybrid star sequences.
Any parametrization of the hybrid model, which varies only by the onset density of the second phase, will cross this point.
This is illustrated in figure~\ref{fig-1}.
\begin{figure}[h]
\centering
\includegraphics[width=8cm]{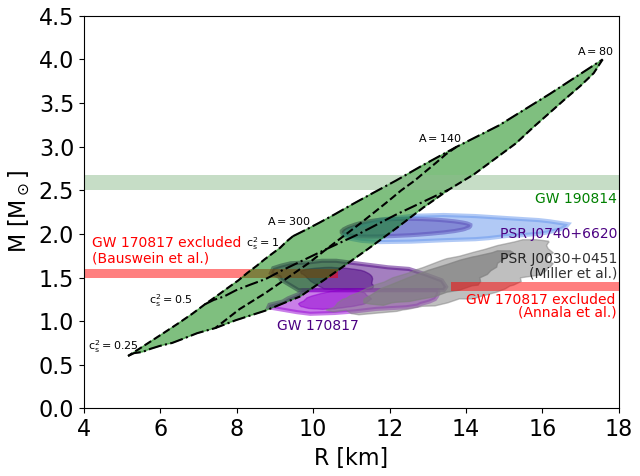}
\caption{
Location of the special point (green area) for $0.25 < c_s^2 < 1.0$ and $80 < A [{\rm MeV/fm}^3] < 300$, lines of constant $c_s^2$ and $A$ are shown for comparison. 
Figure adopted from \cite{Cierniak:2021knt}. 
Observational data taken from \cite{Miller:2021qha,Miller:2019cac,Riley:2021pdl,Riley:2019yda,LIGOScientific:2020zkf,Bauswein:2017vtn,Annala:2017llu}}
\label{fig-1a}
\end{figure}

As was shown in \cite{Cierniak:2020eyh,Blaschke:2020vuy,Cierniak:2021knt}, this feature can be used to characterize a broad family of EoS, without the need to fully scan the parameter space of the model, and to easily compare model predictions with observational data. Most recently, this has been demonstrated in \cite{Cierniak:2021knt}, where a prediction on the radius range of a hybrid star fulfilling PSR J0740+6620 mass--radius measurement was derived.

A detailed map of the special point locations and their dependence on the CSS model parameters has been made in \cite{Cierniak:2021knt}, shown here as figure~\ref{fig-1a}, and will be used in the following chapter.
It has been shown in \cite{Cierniak:2020eyh,Cierniak:2021knt} that this map is independent of the underlying hadronic EoS, the phase transition density and the type of the phase transition (first order, crossover etc.). 
Therefore, it can be used to isolate the hybrid neutron star branch and analyze observational data in the context of constraints on the parameters of the  microscopic model for deconfined quark matter.

\section{Parameter fit}
\label{sec-3}

\begin{figure}[h]
\centering
\includegraphics[width=5.5cm]{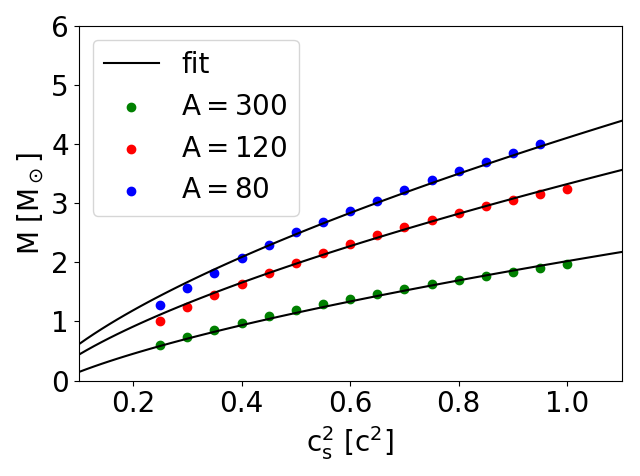}
\hspace{0.5cm}
\includegraphics[width=5.5cm]{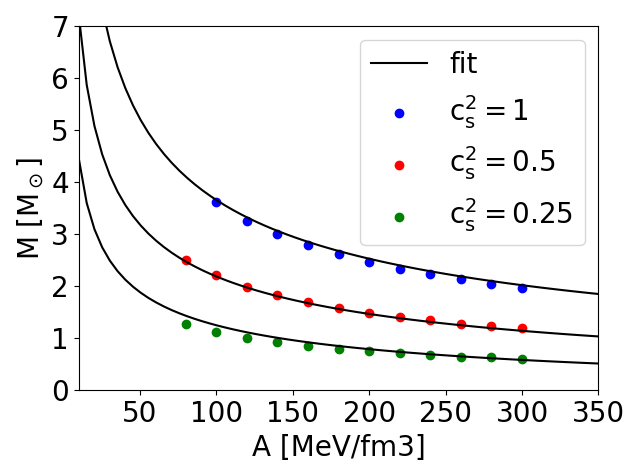}
\caption{Masses of the special point location (colored points) as function of the CSS parameters $c^2_s$ (left panel) and $A$
(right panel) compared to the fit formula~(\ref{eq-fit}) (black line). See text for details.}
\label{fig-2}
\end{figure}

\begin{figure}[h]
\centering
\includegraphics[width=5.5cm]{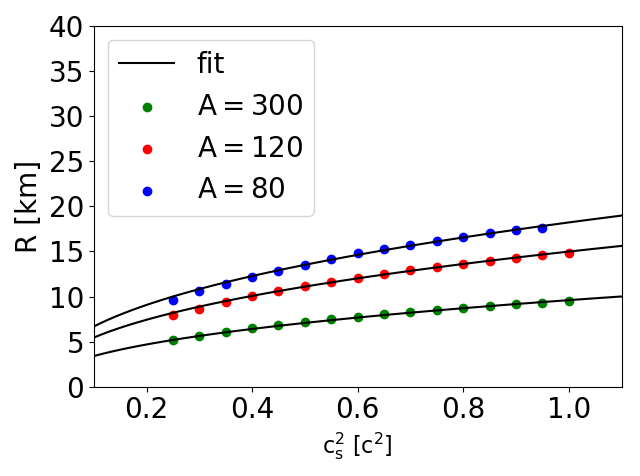}
\hspace{0.5cm}
\includegraphics[width=5.5cm]{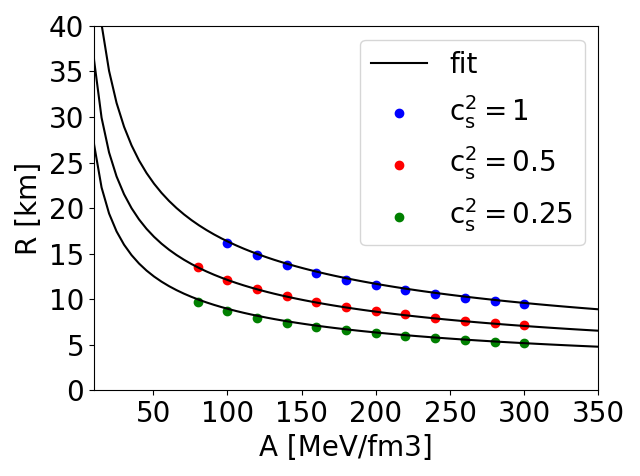}
\caption{Radii of the special point location (colored points) as function of the CSS parameters $c^2_s$ (left panel) and $A$
(right panel) compared to the fit formula~(\ref{eq-fit}) (black line). See text for details.}
\label{fig-3}
\end{figure}

The coordinates of the special point are to a good approximation described by the following relationship
\begin{equation} 
\label{eq-fit}
    \begin{cases}
        R_{SP}[{\rm km}] = a_R \cdot \left(c^2_s\right)^{b_R} \cdot \left(A[{\rm MeV/fm}^3]\right)^{c_R} + d_R,\\
        M_{SP}[M_\odot] = a_M \cdot \left(c^2_s\right)^{b_M} \cdot \left(A[{\rm MeV/fm}^3]\right)^{c_M} + d_M.
    \end{cases}
\end{equation}
The accuracy of this approximation can be judged from figures~\ref{fig-2} and~\ref{fig-3}, where 
the data points were taken from \cite{Cierniak:2021knt} (see also figure~\ref{fig-1a}) and the black solid lines are obtained using the fit formulas of Eq.~(\ref{eq-fit}) with the parameters listed in table~\ref{tab-1}.

\begin{figure}[h]
\centering
\includegraphics[width=5.5cm]{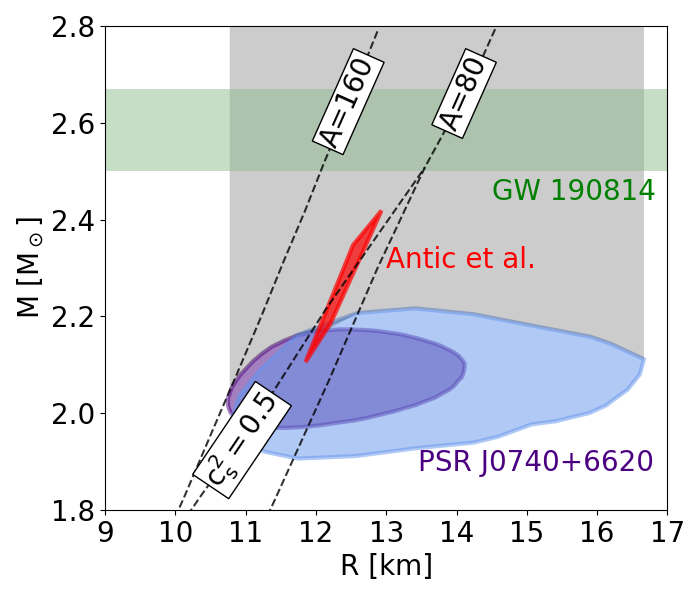}
\hspace{0.5cm}
\includegraphics[width=5.5cm]{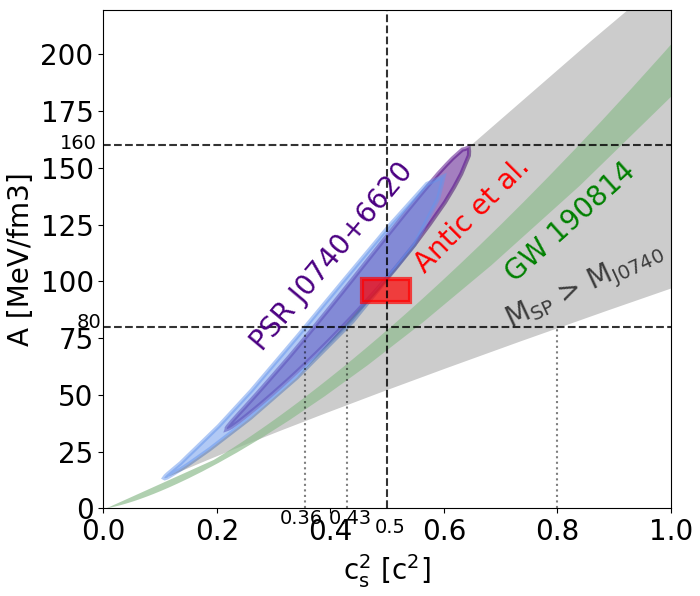}
\caption{Left panel: Neutron star mass--radius diagram with the 1$\sigma$ results of the NICER measurement of PSR J0740+6620 \cite{Miller:2021qha,Riley:2021pdl}. The light green band is the mass range for the lighter object in the binary merger GW190814.
The grey area approximates the region of special point locations for which the hybrid star sequence would fulfill the radius constraint from PSR J0740+6620. For orientation, we show limiting values of constant $A$ as well as $c_s^2=0.5$. 
In the small red area lie all physical parametrizations of the color superconducting nonlocal NJL model of quark matter   
\cite{Antic:2021zbn}
Right panel: The same areas and lines as in the left panel, translated to the coordinate system spanned by the CSS model parameters $c^2_s$ and $A$. 
}
\label{fig-4}
\end{figure}

\begin{table}
\centering
\caption{Parameters of the fit formulas Eq.~(\ref{eq-fit}) for the special point location in the $M-R$ diagram.}
\label{tab-1}
\begin{tabular}{l|llll}
\hline
X & $a_X$ & $b_X$ & $c_X$ & $d_X$  \\\hline
R & 145.938 & 0.420 & -0.47 & -0.4 \\
M & 35.332 & 0.648 & -0.47 & -0.4 \\\hline
\end{tabular}
\end{table}
By inverting the fit (\ref{eq-fit}) we were able to translate existing observational constraints on neutron star masses and radii to constraints on the CSS model parameter space. This is the main result of this work, shown in Fig.~\ref{fig-4}.

\section{Conclusions}
\label{sec-4}

The special point typically corresponds to a mass slightly below $M_{\rm max}$, no less than $M_{\rm max}-M_{\rm SP}=0.2~M_{\odot}$ \cite{Cierniak:2020eyh}.
If we assume $M_{\rm SP} \approx M_{\rm max}$ and that PSR J0740+6620 represents the heaviest possible neutron star, we arrive at a narrow band of CSS model parameters that remain in agreement with this observation.
Further demanding the $A$ parameter not  to fall below a value of 80 MeV/fm$^3$, a value which ensures a phase transition not below the saturation density (cf. \cite{Cierniak:2021knt}), we can determine a narrow band of possible values for the speed of sound parameter $c^2_s$. For $A=80$ MeV/fm$^3$, the possible range is $0.36 \leq c^2_s \leq 0.43$.
Higher values of $A$ would subsequently drive the speed of sound to greater values with a limit in the vicinity of 
$c^2_s \approx 0.6$.

Translating back the requirement of a transition at supersaturation density, we arrive at a limit represented by the black dashed line in the left panel of Fig.~\ref{fig-4}. Only the left side of this limit can be achieved with the CSS model assuming 
$M_{\rm max} \approx M_{J0740}$.
A potential revision of the measurement, that would suggest a higher radius for the star would signal, that this is not the highest possible mass for a neutron star, or that the CSS model does not accurately approximate the cold and dense matter EoS.

Finally, the grey band of the right panel of figure~\ref{fig-4} shows the available parameter space of the CSS model, assuming $M_{\rm max}>M_{J0740}$, for which the model would remain in agreement with the NICER observation. 
For $A=80$ MeV/fm$^3$, this results in an upper limit on $c^2_{s,max}=0.8$, along with the previous lower limit of $c^2_{s,min}=0.36$.

We have demonstrated how the special point feature of hybrid neutron stars can be applied to provide direct limits on the chosen quark matter model parameters by translating the novel NICER measurement of PSR J0740+6620 to the parameter space of the CSS model.
Further constraints can be derived from existing measurements, a task which we leave for subsequent studies. Due to the link between the phenomenological CSS model and the color--superconducting nlNJL model illustrated in \cite{Antic:2021zbn}, such limits can provide valuable insight into the phenomenon of color--superconductivity in quark matter at high densities.

\section*{Acknowledgements}

M.C. and D.B. acknowledge support from the Polish National Science Centre (NCN) under grant number 2019/33/B/ST9/03059. The work of D.B. was supported by the Russian Foundation for Basic Research under grant No. 18-02-40137 and by the Federal Program "Priority-2030".

\bibliography{main}

\end{document}